\documentclass[preprint,aps,prb]{revtex4}

\usepackage{graphicx}
\usepackage{dcolumn}
\usepackage{bm}

\begin{document}
\title{Indirect exchange in GaMnAs bilayers via spin-polarized
inhomogeneous hole gas: Monte Carlo simulation}
\author{M. A. Boselli, I. C. da Cunha Lima}
\address{Instituto de F\'\i sica, Universidade do Estado do Rio de Janeiro\\
Rua S\~{a}o Francisco Xavier 524, 20.500-013 Rio de Janeiro, R.J., Brazil}
\author{A. Ghazali}
\address{Groupe de Physique des Solides, Universit\'{e}s Paris 6 et Paris 7\\
Tour 23, 2 Place Jussieu, F-75 251 Paris Cedex 05, France}

\date{\today}

\begin{abstract}
The magnetic order resulting from an  indirect exchange between
magnetic moments provided by spin-polarized hole gas in the
metallic phase of a Ga$_{1-x}$Mn$_{x}$As double layer structure is
studied \textit{via} Monte Carlo simulation. The coupling
mechanism involves a perturbative calculation in second order of
the interaction between the magnetic moments and carriers (holes).
We take into account a possible polarization of the hole gas due
to the existence of an average magnetization in the magnetic
layers, establishing, in this way, a self-consistency between the
magnetic order and  the electronic structure. That interaction
leads to an internal ferromagnetic order inside each layer, and a
parallel arrangement between their magnetizations, even in the
case of thin layers. This fact is analyzed in terms of the inter-
and intra-layer interactions.

\end{abstract}

\maketitle
\section{Introduction}

\label{intro}

During the last decade a new interest arose in the study of the
magnetic order in layered materials. This area includes the study
of magnetic semiconductor pseudo-binary alloys like
A$_{1-x}$M$_{x}$B, where M stands for a magnetic ion. These alloys
are called Diluted Magnetic Semiconductors (DMS). \cite{dms1,dms2}
A few years ago some groups
\cite{van1,van2,oiwa1,oiwa2,matsukura,ohno1,ohno2} succeeded in
producing homogeneous samples of Ga$_{1-x}$Mn$_{x}$As alloys with
$x$ up to $7\%$ using low temperature ($200-300^{o}$ C) Molecular
Beam Epitaxy (MBE) techniques. Mn is a transition metal having its
$3d$ level half filled with five electrons, in such a way that it
carries a spin of $5\hbar /2$, according to the Hund's rule. In
Ga$_{1-x}$Mn$_{x}$As alloys a substitutional Mn acts as an
acceptor (it binds one hole), and at the same time it carries a
localized magnetic moment, due to its \textit{3d} shell. For
$x=0.053$, the alloy is a metallic ferromagnet, \cite{matsukura}
the Curie-Weiss temperature is 110K, and the free hole
concentration is near $1-2\times 10^{20}$ cm$^{-3}$. The
ferromagnetic order in the metallic phase is understood as
resulting from the indirect exchange between the Mn ions due to
the local spin polarization in the hole gas. \cite{bose01} At
present a vast literature exists about different possible
mechanisms for that indirect exchange.
\cite{dietl,bose01,macdon,bhatt} The free hole concentration in
the metallic phase is a fraction (10-20$\%$) of the total
concentration of Mn. This is understood as due to the presence of
As anti-sites and interstitial Mn.  The magnetic ordering
resulting from indirect exchange {\it via} spin-polarized free
carriers implies in a spin coherence length  larger than the
average distance between localized magnetic moments. The
possibility of having a DMS based on GaAs opens a wide range of
potential applications as, for instance, in integrated
magneto-optoelectronic devices. \cite{nat}

In this work we extended a confinement-adapted
Ruderman-Kittel-Kasuya-Yosida  (RKKY) \cite{bose01} mechanism  to
study the magnetic order resulting from the indirect exchange
between magnetic moments in a GaAs/Ga$_{1-x}$Mn$_{x}$As
nanostructure with two DMS layers. A temperature dependent Monte
Carlo (MC) simulation is performed to determine the resulting
magnetic phases. This article is organized as follows. In
Sec.~\ref{spin-rkky} we present the calculation of the indirect
exchange for a confined spin-polarized Fermi gas in a
semiconductor heterostructure. In Sec.~\ref{elec} the
self-consistent calculation of the heavy-hole single band
electronic structure is presented in some detail. In
Sec.~\ref{moncar} the Monte Carlo simulation is performed to
determine the resulting magnetic phases and the relevant
properties. In Sec.~\ref{comments} we present our conclusions. Our
calculations reveal that a ferromagnetic phase exists even in thin
layers in the structure containing two DMS layers. Considering the
intra-layer and the inter-layer interactions as two independent
mechanisms, we conclude that the intra-layer interaction is
dominant in what concerns the internal magnetic order of each
layer, while the inter-layer interaction alone, which determines
the relative orientation of the two average magnetizations, is
responsible for a transition temperature four times lower than
that resulting from the two mechanism acting together. The roles
of the carrier concentration and of the cutoff parameter are also
analyzed.

\section{Indirect exchange via spin-polarized Fermi gas}

\label{spin-rkky}

The indirect exchange between localized magnetic moments in
quasi-two-dimensional structures  mediated by a Fermi gas has been
addressed several
times.\cite{koren,rkky4,larsen,beal,utiv,aristov,balten1,balten2}
Basically, it deals with a confined electron (or hole) gas being
locally spin-polarized by magnetic moments distributed in a layer.

The interaction potential between the Fermi gas and the set of
localized magnetic moments is well described by the Kondo-like
exchange term:

\begin{equation}
H_{\text{ex}}=-I\sum_i\vec S_i\cdot \vec s(\vec r)\delta(\vec r-
\vec R_i), \label{hund}
\end{equation}
where the localized spin of the Mn ion $\vec S_i$ at position
$\vec R_i$ will be treated as a classical variable, and $\vec
s(\vec r) $ is the spin operator of the carrier at position $\vec
r$; $I$ is the $sp-d$ interaction.

If $\hat \psi _\sigma (\vec r)$ and $\hat \psi _\sigma ^{\dagger }
(\vec r)$ describe the fermion field operator for spin $\sigma $,
then

\begin{equation}
s^z(\vec r)=\frac 12(\hat \psi _{\uparrow }^{\dagger }(\vec r)\hat
\psi _{\uparrow }(\vec r)-\hat \psi _{\downarrow }^{\dagger }(\vec
r)\hat \psi _{\downarrow }(\vec r)),  \label{sigz}
\end{equation}

\begin{equation}
s^{+}(\vec r)=\hat \psi _{\uparrow }^{\dagger }(\vec r)\hat \psi
_{\downarrow }(\vec r),  \label{sig+}
\end{equation}

\begin{equation}
s^{-}(\vec{r})=\hat{\psi}_{\downarrow }^{\dagger }(\vec{r})
\hat{\psi}_{\uparrow }(\vec{r}),  \label{sig-}
\end{equation}
with the usual definitions of $s^{+}=s_{x}+is_{y}$, and
$s^{-}=s_{x}-is_{y}$. In terms of field operators, the (Kondo)
exchange term in the Hamiltonian is
\begin{equation}
H_{\text{ex}}=-\frac{I}{2}\sum_{i}\{S_{i}^{z}[\psi_{\uparrow}^{\dagger}(\vec{r}_i)
\psi_{\uparrow}(\vec{r}_i)-\psi_{\downarrow}^{\dagger}(\vec{r}_i)
\psi_{\downarrow}(\vec{r}_i)]+
S_{i}^{+}\psi_{\downarrow}^{\dagger}(\vec{r}_i)\psi_{\uparrow}(\vec{r}_i)+
S_{i}^{-}\psi_{\uparrow}^{\dagger}(\vec{r}_i)\psi_{\downarrow}(\vec{r}_i)\}.
\label{hexfield}
\end{equation}

We consider  holes in a semiconductor heterostructure that are
confined in the growth direction, assumed to be the z-axis. The
total Hamiltonian, $H=H_0+H_{\text{ex}}$ includes  in $H_0$ the
kinetic part, the confinement potential and the Hartree as well as
exchange and correlation terms. \cite{luc} Neglecting scattering
from impurities, holes in the effective mass approximation are
free particles in the ($x,y$) plane, i.e., in the plane parallel
to the layer interfaces. Their field operator can be written as

\begin{equation}
\hat{\psi}_{\sigma }(\vec{r})=\frac{1}{\sqrt{A}}\sum_{n,\vec{k}}e^{i\vec{k}.%
\vec{\rho}}\phi _{n,\sigma}(z)\eta_{\sigma} c_{n,\vec{k},\sigma },
\label{field}
\end{equation}
where $A$ is the normalization area, $\vec{k}$ is a wave vector in
the ($x,y$) plane, $\eta _{\sigma}$ is the spin tensor for the
polarization $\sigma$, $\phi _{n,\sigma}(z)$ is the envelope
function which describes the motion of the fermion in the
$z$-direction, and $c_{n,\vec{k},\sigma }$ is the fermion
annihilation operator for the state ($n, \vec{k},\sigma $).
$\vec{\rho}$ represents a vector in the ($x,y$) plane.

The confined RKKY indirect exchange is a second-order perturbative
treatment. It gives two terms for the correction to the ground
state energy of the system formed by the set of (classical)
localized moments and the Fermi gas. \cite{bose01} The first one
is the self-energy term, which corrects the site energy. The
second is the interaction between the localized magnetic moments,
the RKKY exchange, allowing the interaction to be written as a
Heisenberg Hamiltonian:
\begin{equation}
\delta E^{(2)}= -\sum_{i,j}J_{ij}\textbf{S}_i\cdot \textbf{S}_j.
\label{unpoljij}
\end{equation}
The RKKY approach can be extended to spin-polarized states. This
is, in one sense, going beyond the second-order perturbation. In
that case no simple spin-spin scalar product is obtained, as in
Eq.~(\ref{unpoljij}), since the spin polarization breaks the
rotational symmetry establishing a preferential direction, namely
that of the average magnetization.

In the second-order perturbation, the correction to the energy
which is bilinear in the magnetic ion dipole moment is obtained
from
\begin{equation}
\delta E^{(2)}=-\sum_{t',F'}
\frac{<F,t|H_{\text{ex}}|F',t'><F',t',|H_{\text{ex}}|F,t>}
{E^{0}_{F',t'}-E^{0}_{F,t}}. \label{basic}
\end{equation}
Here, the state $|F,t>$ is understood as the direct product of the
state of the system of classical localized moments and that of the
quantum spin-polarized gas. Substituting Eq.~(\ref{hexfield}) into
Eq.~(\ref{basic}) we obtain, after a lengthy calculation, the
effective Hamiltonian which includes explicitly the spin-flip
terms:
\begin{equation}
H_{eff}=-\sum_{i,j}
(C_{ij}^{\uparrow\uparrow}+C_{ij}^{\downarrow\downarrow})S_i^zS_j^z+
(C_{ij}^{\uparrow\downarrow}+C_{ij}^{\downarrow\uparrow})(S_i^xS_j^x+S_i^yS_j^y),
\label{newrkky}
\end{equation}
where the spin-flip term is given by:
\begin{eqnarray}
C_{ij}^{\uparrow\downarrow}=-\sum_{n\in\uparrow}\sum_{n'\in\downarrow}\sum_{\textbf{k}\textbf{q}}
(\frac{I}{2A})^2\frac{1}{\epsilon_{n',\textbf{k}+\textbf{q}}-\epsilon_{n,\textbf{k}}}
\exp [i\textbf{q}(\textbf{R}_j-\textbf{R}_i)]\times \nonumber \\
\phi^*_{n'}(z_i)\phi_{n}(z_i)\phi^*_{n}(z_j)\phi_{n'}(z_j)\theta(E_F-\epsilon_{n,\textbf{k}})
\theta(\epsilon_{n',\textbf{k}+\textbf{q}}-E_F). \label{defcij}
\end{eqnarray}
Notice that the arrows refer to spins up or down for holes, not
the orientation of the localized magnetic moments at the Mn sites.
Similar definitions hold for the other $C_{ij}^{\sigma\sigma '}$.
Eq.~(\ref{defcij}) can be expressed in terms of the Lindhard
function: \cite{abrik,keld}
\begin{equation}
\chi ^{n,n^{\prime }}(\vec q)=\sum_{\vec k} \frac{\theta
(E_F-\epsilon _{n,\vec k})-\theta (E_F-\epsilon _{n^{\prime },
\vec k+\vec q})}{\epsilon _{n^{\prime },\vec k+\vec q}-\epsilon
_{n,\vec k}},  \label{modlin}
\end{equation}
and its real-space Fourier transform:
\begin{equation}
\chi^{n,n^{\prime}}(\textbf{R}_{ij})=\sum_{\vec q}\exp [-i\vec
q\cdot \textbf{R}_{ij}]\chi ^{n,n^{\prime }}(\vec q).
\end{equation}
Substituting Eq.~(\ref{modlin}) in Eq.~(\ref{defcij}) we obtain
\begin{equation}
C_{ij}^{\mu\nu}=-\sum_{n\in\mu}\sum_{n'\in\nu}\ (\frac{I}{2A})^2
\phi^*_{n'}(z_i)\phi_{n}(z_i)\phi^*_{n}(z_j)\phi_{n'}(z_j)
\chi^{n,n^{\prime}}(\textbf{R}_{ij}), \label{cijmn}
\end{equation}
{\it where the summations on $n$ and $n\prime$ are restricted to
those sub-bands with the proper spin polarization}.

The intra-subband real space Lindhard function
$\chi^{n,n}(\textbf{R}_{ij}) $ has been derived by several
authors: \cite{bose01,koren,utiv,beal,aristov}

\begin{equation}
\chi ^{n,n}(R_{ij})=-\frac{m_{t}^{\ast }A^{2}}{4\pi \hbar ^{2}}%
k_{F}^{(n)2}[J_{0}(k_{F}^{(n)}R_{ij})N_{0}(k_{F}^{(n)}R_{ij})+J_{1}(k_{F}^{(n)}R_{ij})N_{1}(k_{F}^{(n)}R_{ij})],
\label{chiintra}
\end{equation}
where $m_{t}^{\ast }$ is the transversal (parallel to the interfaces) effective mass, and
$k_{F}^{(n)}$ refers to the Fermi wavevector of subband $n$.

The inter-subband terms cannot be expressed in a closed form.
\cite{bose01} They must be obtained numerically from the integral
\begin{equation}
\chi^{n,n^{\prime}}(\textbf{R}_{ij})=
-\int_{0}^{\infty
}dqq F_{n,n^{\prime }}(q)J_{0}(qR_{ij}),  \label{stj}
\end{equation}
where
\begin{equation}
F_{n,n^{\prime }}(q)=\frac{A^2m_{t}^{\ast }}{8\pi^2\hbar ^{2}}(1-\frac{\Delta
_{n^{\prime },n}}{q^{2}})[1-\sqrt{1-(\frac{2k_{F}^{(n)}q}{q^{2}+\Delta
_{n^{\prime },n}})^{2}}\theta (q^{2}+\Delta _{n^{\prime
},n}-2qk_{F}^{(n)})]\theta (\epsilon _{n^{\prime}}-E_{F}),
\end{equation}
with $\Delta _{n^{\prime },n}=2m_{t}^{\ast }\cdot (E_{n^{\prime
}}-E_{n})/\hbar ^{2}$.

\section{Spin-polarized electronic structure for magnetic
multilayers} \label{elec}

 In order to obtain the spin-polarized electronic
structure for holes  given an average magnetization of the Mn
ions, we solve self-consistently the heavy hole single band
Schr\"odinger equation in the reciprocal space. The hole system is
supposed to be homogeneous in the $xy$ plane, so the Hartree term
depends only on the coordinate $z$, $U_{H}(\vec{r})=U_{H}(z).$ For
the purpose of obtaining the electronic structure we treat the
magnetic interaction as being due to an uniform magnetization in
the DMS layers. If a net magnetization exists, it will polarize
the hole gas. This problem is solved self-consistently by a
secular matrix equation in the reciprocal space. The method would
be exact were not for cutting the matrix size. The advantage is
that it provides spin-polarized eigenvalues and eigenfunctions
with high accuracy, not only for bound states, but also for a high
number of scattering states. For each spin, we define the
wavefunction Fourier Transform (FT):

\begin{equation}
\psi _{\sigma }(\vec{r})=\int d^{3}q\exp (i\vec{q}.\vec{r}) \psi
_{\sigma } (\vec{q}).  \label{psiq}
\end{equation}
The hole eigenstates will be obtained by discretizing the
integrals on $\vec{q}$ appearing in
\begin{equation}
\int d^{3}r\psi _{\sigma }^{\ast }(\vec{r})(H-E)\psi _{\sigma
}(\vec{r})=0. \label{expect}
\end{equation}
When integrating the magnetic term in the Hamiltonian over
$\vec{r}$, we assumed the magnetic impurities to be uniformly
distributed in each one of the Ga$_{1-x}$Mn$_{x}$As DMS layers,
all of them having the same thermal average magnetization. This
treatment includes not only the ferromagnetic phase but also
phases where a partial magnetization is observed, as the
``canted-spin'' phases. \cite{bose01,bose02} Therefore, taking an
homogenous concentration of $N_i$ magnetic impurities in the
$j$-th layer, we have:
\begin{eqnarray}
-I\int d^{3}r\exp [i(\vec{q}-\vec{q}^{\prime
}).\vec{r}]\sum_j\sum_{i\in j}^{N_{i}}
\vec{s}(\vec{r}).\vec{S}(\vec{R_{i}})\delta (\vec{r}-\vec{R_{i}})
\simeq  \nonumber \\
-I \frac{\sigma }{2}\sum_j<M>_j\sum_{i\in j}^{N_{i}}\exp
[i(\vec{q}-\vec{q}^{\prime
}). \vec{R_{i}}]=  \nonumber \\
-I\frac{\sigma }{2}\sum_j<M>_j\frac{N_{i}}{V}\int d^{3}R_{i}\exp
[i(\vec{q}-\vec{q} ^{\prime }).\vec{R_{i}}]= \nonumber \\
-N_{0}\beta\frac{\sigma }{2}
x\sum_j<M>_jF^j_{DMS}(q_{z}-q_{z}^{\prime })(2\pi )^{3} \delta ^{2} (\vec{q}%
_{\parallel}-\vec{q}_{\parallel }^{\prime }),  \label{imag}
\end{eqnarray}
where $\sigma =\pm 1$ for spin parallel (upper sign) or
anti-parallel (lower sign) to the magnetization, and
$\frac{N_i}{V}=xN_0$ is the impurity density, in terms of $N_0$,
the density of the cation ions. The {\it sp-d} exchange constant
for holes is usually written as $\beta$. Here we used
$N_{0}\beta=-1.2$eV. \cite{okaba} $F^j_{DMS}$ is the integral
performed on the $z$ -coordinate inside the $j$-th DMS layer:
\begin{equation}
F^j_{DMS}(q)\equiv \frac{1}{2\pi }\int_jdz\exp [iq.z].
\label{form}
\end{equation}

Comparing Eqs.~(\ref{imag}) and (\ref{form}), the latter being
just the Fourier transform of an unit barrier function within the
range of the $j$-th DMS layer, we see that the thermal average
magnetization $<M>_{j}$ polarizes the hole gas by introducing
additional effective confining potentials given by
\begin{equation}
V_{mag}^{eff}(z)=-N_{0}\beta x\frac{\sigma
}{2}\sum_{j}<M>_{j}g_{j}(z), \label{effecmag}
\end{equation}
where $g_{j}(z)=1$ if $z$ lies inside the $j$-th layer, and
$g_{j}(z)=0$ otherwise. Next, we define
\begin{equation}
U_{eff}(q)=\frac{1}{2\pi }\int dz\exp
[iq.z][U_{c}(z)+V_{mag}^{eff}(z)+U_H(z)],  \label{effective}
\end{equation}
where $U_{c}(z)$ and $U_H(z)$ are the confining and the Hartree
potentials, respectively. The  eigenvalues and eigenfunctions at
the bottom of the 2-D subbands may be obtained by solving the
secular equation for each spin polarization:
\begin{equation}
\det \left\{ \left[ \frac{\hbar ^{2}q_{z}^{2}}{2m^{\ast
}}-E\right] \delta (q_{z}-q_{z}^{\prime
})+U_{eff}(q_{z}-q_{z}^{\prime })\right\} =0.  \label{secul}
\end{equation}

\section{Monte Carlo simulation: Interlayer interaction and magnetic ordering}

\label{moncar}

In the present work we have focused our attention on metallic
systems with the Mn concentration $x=5\%$, and we have neglected
possible superexchange (anti-)ferromagnetic interaction between
the nearest neighbors and the next-nearest neighbors pairs. We
have performed extensive Monte Carlo (MC) simulations in order to
determine the possible magnetic order  occurring in a system
containing a double layer of Ga$_{1-x}$Mn$_x$As. Classical spins
${\bf S}_{i}$, representing the localized magnetic moments of the
Mn ions, are randomly distributed on the cation sites with
concentration $x$. They  are assumed to interact through the
exchange Hamiltonian defined by Eq.~(\ref{newrkky}).

The interaction derived in Sec.~\ref{spin-rkky} is assumed to be
effective within a cutoff radius which corresponds to the carrier
spin coherence length.  This cutoff is taken tentatively as
$R_{c}=2.5a$, and $R_{c}=2a$, where $a$ is the fcc lattice
parameter of GaAs. These values correspond to five and four
monolayers (ML), respectively. This choice makes the smallest
value assumed for $R_{c}$ nearly equal to the hole mean free path
estimated from bulk transport measurements. \cite{matsukura}

The electronic structure was calculated for a system of two DMS
layers grown inside a  GaAs quantum well. We have studied two
different DMS layer widths, 10 and 20 \AA. The GaAs spacer
separating the DMS layers has been considered as having widths
varying from 5 \AA\  to 60 \AA. The whole system consists of a
200\AA\ quantum well with infinite barriers in the z-direction, in
the center of which the  two GaMnAs layers and a GaAs spacer are
disposed, as shown in Fig.~{\ref{fig01}. The eigenstates, obtained
according to Sec.~\ref{elec}, were calculated for net
magnetizations $\frac{<M>}{5\hbar/2} =0.0, 0.1, 0.2, \dots 1.0$,
for each one of the  geometries used. Notice that $<M>$ determines
the effective magnetic potential in Eq.~(\ref{effecmag}) and, in
consequence, the whole electronic structure. A typical result for
the electric charge density, and the spin polarization densities
is shown in Fig.~1c. We see that holes are attracted to the region
of the DMS layers by the negatively charged Mn ions. Because
$N_0\beta$ is negative, however, the $sp-d$ interaction (taken
into account as the effective magnetic potential) favors the
occurrence of anti-parallel spin subbands at lower energies.
Several subbands are occupied, due to the high concentration of
free carriers, most of them corresponding to the anti-parallel
spin polarization. Therefore, the total interaction, i.e., the
confinement, the Coulomb interaction (Hartree and correlation
terms) plus the $sp-d$ interaction results in an inhomogeneous
distribution of spin polarization and spin charge density. Holes
with parallel spins, at lower density than that of parallel spins,
concentrate on the DMS layers, while the latter, at higher
density, spreads in a much wider region. The inhomogeneity of the
spin polarization density influences the indirect exchange between
localized magnetic moments, and this fact is taken into account
explicitly in the present calculation. For each value of $<M>$ the
exchange interaction terms in Eq.~(\ref{newrkky}) are calculated
and tabulated for  all possible magnetic ions distances $R_{ij}$
in the $(x,y)$ plane according to the zincblende structure within
the cutoff radius.

The MC calculation is performed in a finite box, whose axes are
parallel to the [100] directions. Its dimensions are
$L_{x}=L_{y}$, and $L_{z}=2L$, where $L$ is the width of each DMS
layer. Periodic boundary conditions are imposed in the $(x,y)$
plane. The lateral dimensions are adjusted in such a way that the
total number $N_{s}$ of Mn sites is about 4000, for all samples
with different $L_{z}$. The initial spin orientations of the Mn
ions are randomly assigned in the two DMS layers which are
separated by a distance $d$, the width of the GaAs spacer. At a
given temperature, the energy of the system due to exchange
interaction is calculated, and the equilibrium state for a given
temperature is sought by changing the individual vector spin
orientation according to the Metropolis algorithm. \cite{diep} A
slow cooling stepwise process is accomplished, starting from above
the transition temperature $T_c$, and  making sure that the
thermal equilibrium is reached at every temperature. Every time
the net magnetization increases to reach a value of $n \times
0.1$, $n$ an integer varying form 0 to 10, the tabulated values of
the exchange interaction terms are substituted by those
corresponding to the new reference $<M>$. The tabulated values of
the exchange interaction terms were calculated  from the hole
states (eigenfunctions, eigenvalues and Fermi wavevectors)
resulting from self-consistent calculation described in
Sec.~\ref{elec} for this specific thermal average magnetization
$<M>$. Then, the resulting spin configuration for the Mn ions is
taken as the starting configuration for the next step at a lower
temperature. The Monte Carlo procedure adopted in this way takes
account of the changes on the hole gas polarization due to the
presence of the established  order among the localized magnetic
moments in the Mn ions. Thus the magnetization is determined by a
fully self-consistent process, but the effects of the fluctuations
of the magnetization on the electronic structure are neglected in
this process.

For each temperature, the thermal average magnetization $<M>$, and
the Edwards-Anderson (EA) order parameter $q$ are calculated.
\cite{diep} The latter is defined as
\begin{equation}
q=\frac{1}{N}\sum_{i=1}^{N} \left( \sum_\alpha \left| \frac{1}{t}
\sum_{t^{\prime }=t_{0}}^{t_{0}+t}{S}_{i \alpha}(t^{\prime }) \right|^2
\right)^{1/2},  \label{eadef}
\end{equation}
where $\alpha=x$, $y$ and $z$. In order to avoid spurious results
on the thermal average values of $q$ over a large time interval
$t$, a summation on $ t^{\prime }$ is performed starting from a
time $t_{0}$, when the system already reached the thermal
equilibrium. This is true for other quantities such as energy,
magnetic susceptibility, etc.

The MC simulations have been performed for several samples as
indicated on Table I. The ratio between the carrier concentration
and the Mn concentration is the same for the two DMS layers, and
is chosen to reproduce hole concentrations which in bulk would be
$p = 1 \times 10^{20}$ cm$^{-3}$, and $p = 2 \times 10^{20}$
cm$^{-3}$. In Fig.~\ref{fig02} the normalized magnetization as a
function of temperature is shown for samples \#1 to \#6, with the
DMS layer width of 10 \AA,  bulk hole concentration $p = 1 \times
10^{20}$ cm$^{-3}$, and $R_c = 5$ monolayers (ML). All these
samples show a paramagnetic (P) to a ferromagnetic (FM) phase
transition. The transition temperatures decrease with the increase
of the spacer thickness, pointing to the importance of the
inter-layer interaction. In Fig.~\ref{fig03} all parameters are
kept the same as in the previous figure, but $p = 2 \times
10^{20}$ cm$^{-3}$. The increase on the hole concentration has two
main influences on the magnetization curves: (i) an increase of
$T_c$ and (ii) the possibility of the occurrence of a canted spin
phase (C), with a partial ferromagnetic  order.\cite{bose01} This
canted spin phase can be explained on the basis of the oscillatory
behavior of the exchange interaction. The increase on the hole
concentration produces an increase of the Fermi wave number, and
anti-ferromagnetic (AF) interactions are likely to occur,
according to the oscillatory nature of the (confined) indirect
interaction. The Curie temperatures $T_c$ for samples \# 01 to \#
12 are obtained in the range from 20 K to 49 K.

In Fig.~\ref{fig04} the results are shown for two 20 \AA\  DMS
layers. The hole concentration is $p = 2 \times 10^{20}$
cm$^{-3}$, and the cutoff length is 5 ML.  $T_c$ decreases with
the interlayer separation, as before. However, compared to the 10
\AA\ system, a higher $T_c$ is observed in the present case. In
Fig.~\ref{fig05} the cutoff radius used was 4 ML. We observe here
the same behavior as before, but with higher $T_c$, and also with
higher values of $<M>$. The presence of effective AF interactions
in these samples, due to the oscillatory behavior of the exchange
interaction, may produce canted spin phases, rather than FM
phases. This results from competing ferro- and antiferromagnetic
interactions. It is the case of the so-called frustration.
Although these are ordered phases, as can be observed in the EA
parameters on Fig.~\ref{fig06}, the Mn magnetic moments are not
all parallel and the magnetizations at zero temperature are in
most samples partial, lying on the range of 60\% to 100\%. The
Curie temperatures $T_c$ for samples \# 13 to \# 24 are obtained
in the range from 35 to 64 K.

In order to better understand the importance of  the
spin-polarized indirect exchange on the magnetic order and on the
transition temperature of these layered systems, we have compared
our results with a simulation performed with an unpolarized
exchange. This is done by using the tabulated values of the
exchange for $<M>=0$ in the whole cooling stepwise MC simulation.
For this comparison we have chosen a sample with the same
characteristics of sample \# 1, i.e., two 10 \AA \ DMS layer and a
GaAs spacer of 5 \AA. The result is shown in Fig.~\ref{fig07}. The
presence of polarization more than doubles the Curie temperature.
This occurs because the effective magnetic potential resulting
from the finite thermal average magnetization, and the consequent
polarization of the hole gas favor a stronger concentration of
charge and spin on the  DMS layers, as can be seen in
Fig.~\ref{fig01}, providing a stronger exchange than that for the
unpolarized system.

A calculation was performed to estimate how the  interlayer
interaction is affected by the the width of the spacer, and how
this effect appears in the Curie temperature. To this end, the
exchange interaction inside each DMS layer  was ``turned off'' by
making it artificially equal to zero. Thus, only the Mn ions
belonging to two different DMS layers are interactive. The MC
simulations are then performed on the same basis as described
before. The results are presented in Table II, for the two DMS
layers with 10 \AA \ width separated by 5, 10, 15, 20, 40, and 60
\AA, and for two different hole concentrations, $1 \times 10^{20}$
cm$^{-3}$ in Fig.~\ref{fig08} (samples \#19 to \#24), and $2
\times 10 ^{20}$ cm$^{-3}$ in Fig.~\ref{fig09} (samples \#25 to
\#30). These results show that there is an effective interlayer
coupling able to produce a FM phase. The fact that T$_c$ coming
out exclusively from that interaction is about four times lower
than that for the complete interaction reveals that the main
contribution to the magnetic ordering within each layer is due to
de intralayer coupling. The interlayer interaction acts mainly to
establish the ferromagnetic arrangement between the magnetization
of the two layers. Notice, however, that the resulting magnetic
order affects and is affected by the distribution of charge and
spin of the free carriers in the structure.  We have performed an
exponential fitting of the transition temperature as a function of
the spacer thickness for samples with $L=10\AA\ $, taking into
account the complete interaction and also solely with the
interlayer interaction. This is shown in Fig.~\ref{fig10}. Notice
that the exponential fitting works well in both cases, putting
into evidence the importance of the intra-layer interaction. A
similar calculation was performed to explore the dependence of the
Curie temperature on the DMS layer width, for a fixed spacer
thickness ($d=10\AA\ $). The results appear in Fig.~\ref{fig11},
for $L=10,20,40$ and $50\AA$. The Curie temperature increases with
the DMS width  approaching  the fully filled $T_c$, as a
saturation exponential function.

Up to now the simulations in this work were performed in a single
Monte Carlo sample (for each geometry chosen, given by the set of
parameters $L,d$ and $p$) with a specific distribution of the
sites occupied by the magnetic ions. Next, the calculation is
performed taking into account a configurational average on the
initial distribution of the orientation of individual magnetic
moments $\bf{S}_i$ on the Mn ions, as well as on the sites
$\bf{R}_i$ where these ions are localized. The positions of the Mn
ions identify a MC sample. The parameters are chosen as those of
sample $\#01$ in Table I. The site configurational average is
performed, in each case, over three initial configurations, namely
$a,b$ and $c$ (the MC samples). Although the results do not change
significantly, as it can be seen from Fig.~\ref{fig12} which shows
the average magnetization {\it versus} temperature, it is
important to observe general behaviors which appear in the
process. For instance, Fig.~\ref{fig13} shows the susceptibility
{\it versus} temperature, and we can see that, apart the critical
fluctuations to be expected at the transition temperature, which
is perfectly defined in that figure, the fluctuations are more
important at low temperatures than at high temperatures above
$T_C$. Another interesting behavior appears in Fig.~\ref{fig14},
showing the magnetic specific heat ($C_V$) {\it versus}
temperature. Again, the fluctuations are more important at low
temperatures. Besides, the maximum of $C_V$ occurs at a
temperature which is lower than that of the maximum of the
susceptibility, what is reminiscent of the spin-glass behavior,
and can be understood as a consequence of the competing  FM and AF
interactions characteristics of the indirect exchange.

\section{Discussions and Final Comments}
\label{comments}

This Monte Carlo simulations shows that this system with two thin
GaMnAs layers inside a GaAs matrix presents a ferromagnetic, or at
least a canted magnetic phase ordering. The input parameters,
namely the hole concentration, the DMS layer width, the  DMS layer
separation and the interaction cutoff, all influence the magnetic
phase of these systems. The increase of hole concentration favors
the canted phase due to the intrinsic frustration of the indirect
exchange interaction. In the reverse direction, the elimination of
the AF interactions produces FM samples and an increase of the
T$_c$ as already observed in a previous work. \cite{bose01} The
increase of the DMS layer width will increase the number of
interacting magnetic moments raising also T$_c$. As expected,
T$_c$ will be lowered with the increase on the spacer thickness,
since  as the separation increases the interlayer coupling become
weaker and weaker until it  reaches a limiting value. This limit
is finite due to the confinement by the infinite barriers at the
boundaries of the structure. We have quantified the influence of
this parameter on the magnetization of the system. Despite the
fact that this interlayer interaction is four times weaker than
that of the complete system, it is important to notice that it
plays an important role in the existence of a FM phase in
multilayers, and in digital superlattices.

\acknowledgements This work was supported by CENAPAD-SP (Centro
Nacional de Processamento de Alto Desempenho em S\~ao Paulo)
UNICAMP/FINEP-MCT, CNPq and FAPERJ in Brazil, and by the
CAPES-COFECUB Franco-Brazilian Program.

\newpage
\begin{table}[tbp]
\caption{Sample characteristics: L is the  width of each DMS
layer; d is the GaAs spacer width;  the ratio of the carrier
concentration to the Mn concentration, written in terms of a bulk
hole concentration, $p$ ; $T_c$ is the transition temperature for
phases: FM: ferromagnetic, C: canted spin; $R_c$ is the cutoff
radius of the exchange interaction in number of monolayers.}
\label{tab01}
\begin{tabular}{crccccc}
sample & L (\AA) & d (\AA) & $r$ & $R_c$ & phase & $T_c$ (K)
\\ \hline
\#01 & 10 & 05 & 1 & 5 & FM & 38 \\
\#02 & 10 & 10 & 1 & 5 & FM & 31 \\
\#03 & 10 & 15 & 1 & 5 & FM & 28 \\
\#04 & 10 & 20 & 1 & 5 & FM & 23 \\
\#05 & 10 & 40 & 1 & 5 & FM & 21 \\
\#06 & 10 & 60 & 1 & 5 & FM & 20 \\
\#07 & 10 & 05 & 2 & 5 & C  & 49 \\
\#08 & 10 & 10 & 2 & 5 & FM & 39 \\
\#09 & 10 & 15 & 2 & 5 & FM & 35 \\
\#10 & 10 & 20 & 2 & 5 & FM & 34 \\
\#11 & 10 & 40 & 2 & 5 & FM & 32 \\
\#12 & 10 & 60 & 2 & 5 & FM & 22 \\
\#13 & 20 & 05 & 2 & 5 & C  & 62 \\
\#14 & 20 & 10 & 2 & 5 & C  & 63 \\
\#15 & 20 & 15 & 2 & 5 & C  & 59 \\
\#16 & 20 & 20 & 2 & 5 & C  & 59 \\
\#17 & 20 & 40 & 2 & 5 & C  & 53 \\
\#18 & 20 & 60 & 2 & 5 & C  & 51 \\
\#19 & 20 & 05 & 2 & 4 & FM & 64 \\
\#20 & 20 & 10 & 2 & 4 & FM & 56 \\
\#21 & 20 & 15 & 2 & 4 & FM & 52 \\
\#22 & 20 & 20 & 2 & 4 & FM & 49 \\
\#23 & 10 & 40 & 2 & 4 & FM & 35 \\
\#24 & 10 & 60 & 2 & 4 & FM & 45 \\
\end{tabular}
\end{table}

\begin{table}[tbp]
\caption{Same as table 1, but considering only interlayer
interaction.} \label{tab02}
\begin{tabular}{crccccc}
sample & L (\AA) & d (\AA) & $p$ (10$^{20}$ cm$^{-3}$) & $R_c$ &
phase & $T_c$ (K)
\\ \hline
\#25 & 10 & 05 & 1 & 5 & FM & 13 \\
\#26 & 10 & 10 & 1 & 5 & FM & 09 \\
\#27 & 10 & 15 & 1 & 5 & FM & 08 \\
\#28 & 10 & 20 & 1 & 5 & FM & 06 \\
\#29 & 10 & 40 & 1 & 5 & FM & 06 \\
\#30 & 10 & 60 & 1 & 5 & FM & 06 \\
\#31 & 10 & 05 & 2 & 5 & FM & 15 \\
\#32 & 10 & 10 & 2 & 5 & FM & 08 \\
\#33 & 10 & 15 & 2 & 5 & FM & 08 \\
\#34 & 10 & 20 & 2 & 5 & FM & 08 \\
\#35 & 10 & 40 & 2 & 5 & FM & 08 \\
\#36 & 10 & 60 & 2 & 5 & FM & 07 \\
\end{tabular}
\end{table}

\begin{figure}[tbp]
\caption{(a) Structure model used in the present calculation: two
DMS layers inside a GaAs quantum. (b) The effective magnetic
potential (schematic). (c) Density distribution: charge(solid
line), spin polarization(dashed), anti-parallel spin
(dashed-dotted), and parallel spin (dotted).}
\label{fig01}
\end{figure}

\begin{figure}[tbp]
\caption{Normalized magnetization {\it vs} temperature for samples
\#01 to \#06 indicated in Table \protect{\ref{tab01}}.}
\label{fig02}
\end{figure}

\begin{figure}[tbp]
\caption{Same as Fig 2 for samples \#07 to \#12 indicated in Table
\protect{\ref{tab01}}.}
\label{fig03}
\end{figure}

\begin{figure}[tbp]
\caption{Normalized magnetization {\it vs} temperature for samples
\#13 to \#18 indicated in Table \protect{\ref{tab01}}.}
\label{fig04}
\end{figure}

\begin{figure}[tbp]
\caption{Normalized magnetization {\it vs} temperature for samples
\#19 to \#24 indicated in Table \protect{\ref{tab01}}.}
\label{fig05}
\end{figure}

\begin{figure}[tbp]
\caption{Edward-Anderson order parameter {\it vs} temperature for
samples \#13 to \#18 indicated in Table \protect{\ref{tab01}}.}
\label{fig06}
\end{figure}

\begin{figure}[tbp]
\caption{Comparison for the polarized (solid line) and unpolarized
(doted line) system. The characteristics of the sample are the
same as sample \#01, indicated in Table \protect{\ref{tab01}}.}
\label{fig07}
\end{figure}

\begin{figure}[tbp]
\caption{Normalized magnetization {\it vs} temperature for samples
\#25 to \#30 indicated in Table \protect{\ref{tab02}} taking into
account only the interlayer interaction.} \label{fig08}
\end{figure}

\begin{figure}[tbp]
\caption{Normalized magnetization {\it vs} temperature for samples
\#31 to \#36 indicated in Table \protect{\ref{tab02}} taking into
account only the interlayer interaction.} \label{fig09}
\end{figure}

\begin{figure}[tbp]
\caption{Curie temperature {\it vs} spacer thickness for $L=10\AA\
$, and $p=1.\times 10^{20}$ cm$^{-3}$ for the complete interaction
(samples \#01 to \#06 indicated in Table \protect{\ref{tab01}})
and for the interlayer interaction (samples \#25 to \#30 indicated
in Table \protect{\ref{tab02}}). An exponential fitting
$y(x)=y_0+A \exp(-x/\lambda)$ is shown, with the resulting
parameters: $y_0=20.1$K, $A=31.2$K, $\lambda=9.8\AA\ $ for the
complete interaction; $y_0=5.9$K, $A=15.4$K, $\lambda=6.5\AA\ $
for the interlayer interaction.} \label{fig10}
\end{figure}

\begin{figure}[tbp]
\caption{Curie temperature {\it vs} DMS layer thickness for
$d=10\AA\ $, and $p=2.\times 10^{20}$ cm$^{-3}$  An exponential
fitting $y(x)=y_0(1- \exp(-x/\lambda))$ is shown, with the
resulting parameters: $y_0=85.4$K, and $\lambda=15.2 \AA\ $.}
\label{fig11}
\end{figure}

\begin{figure}[tbp]
\caption{Thermal average magnetization as a function of
temperature for three different magnetic moments configurations
(open symbols), and the corresponding configurational average
(solid square), corresponding to the parameters of sample \#01.}
\label{fig12}
\end{figure}

\begin{figure}[tbp]
\caption{ Magnetic susceptibility {\it vs} temperature, calculated
from thermal fluctuations, for three different magnetic moments
configurations (open symbols), and the corresponding
configurational average (solid square), corresponding to the
parameters of sample \#01.} \label{fig13}
\end{figure}

\begin{figure}[tbp]
\caption{ Magnetic specific heat {\it vs} temperature, calculated
from thermal fluctuations, for three different magnetic moments
configurations (open symbols), and the corresponding
configurational average (solid square), corresponding to the
parameters of sample \#01.} \label{fig14}
\end{figure}

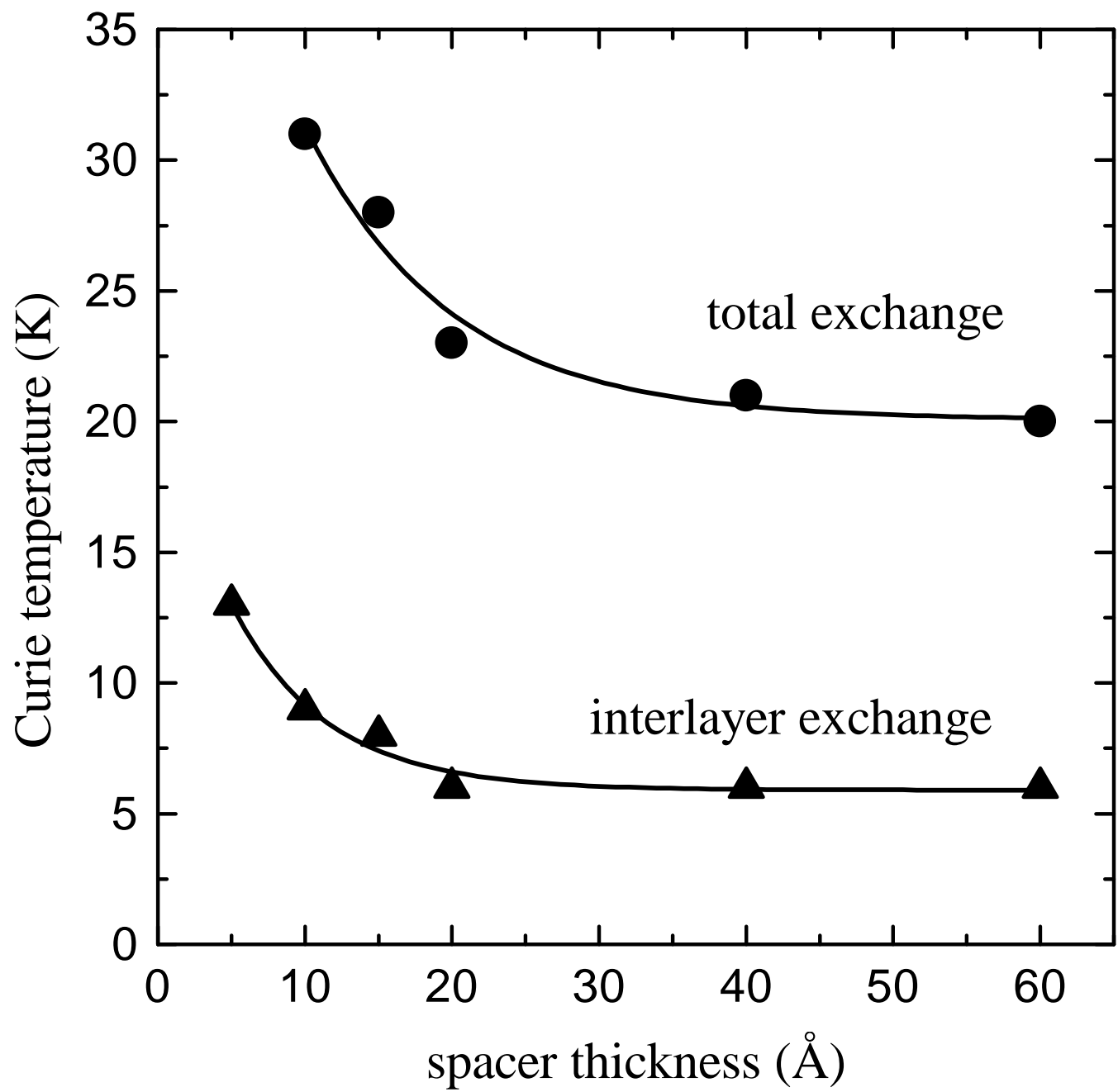
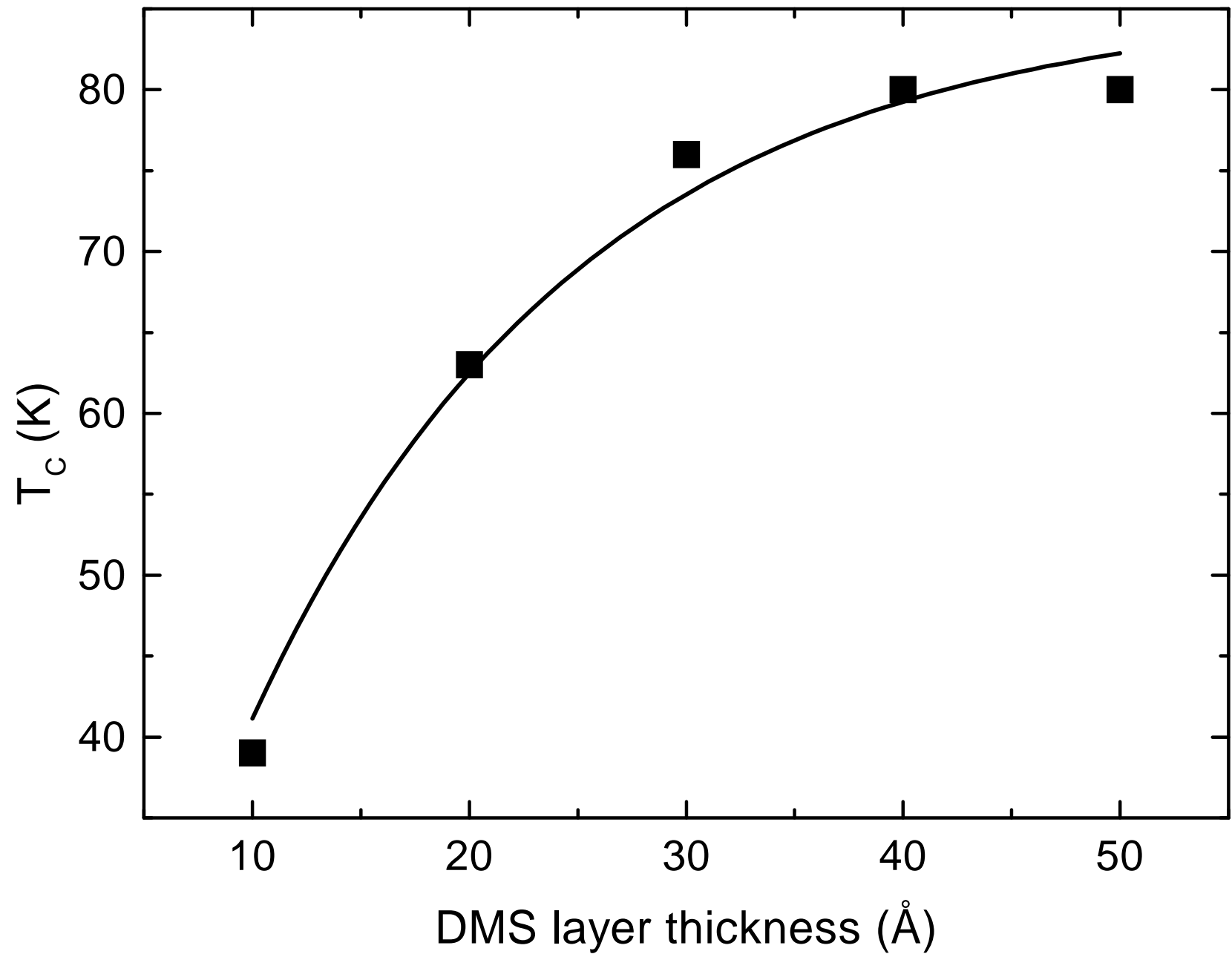
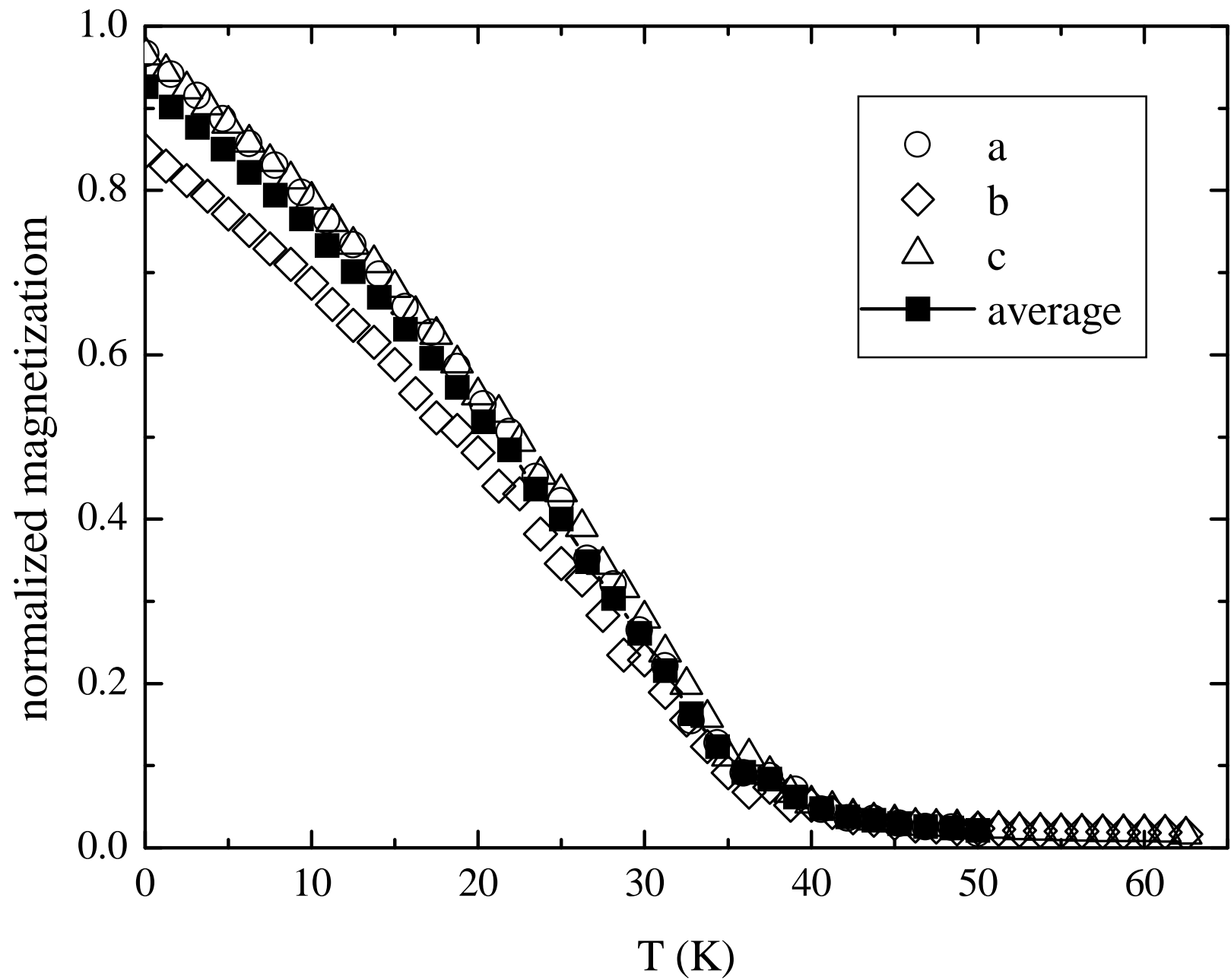
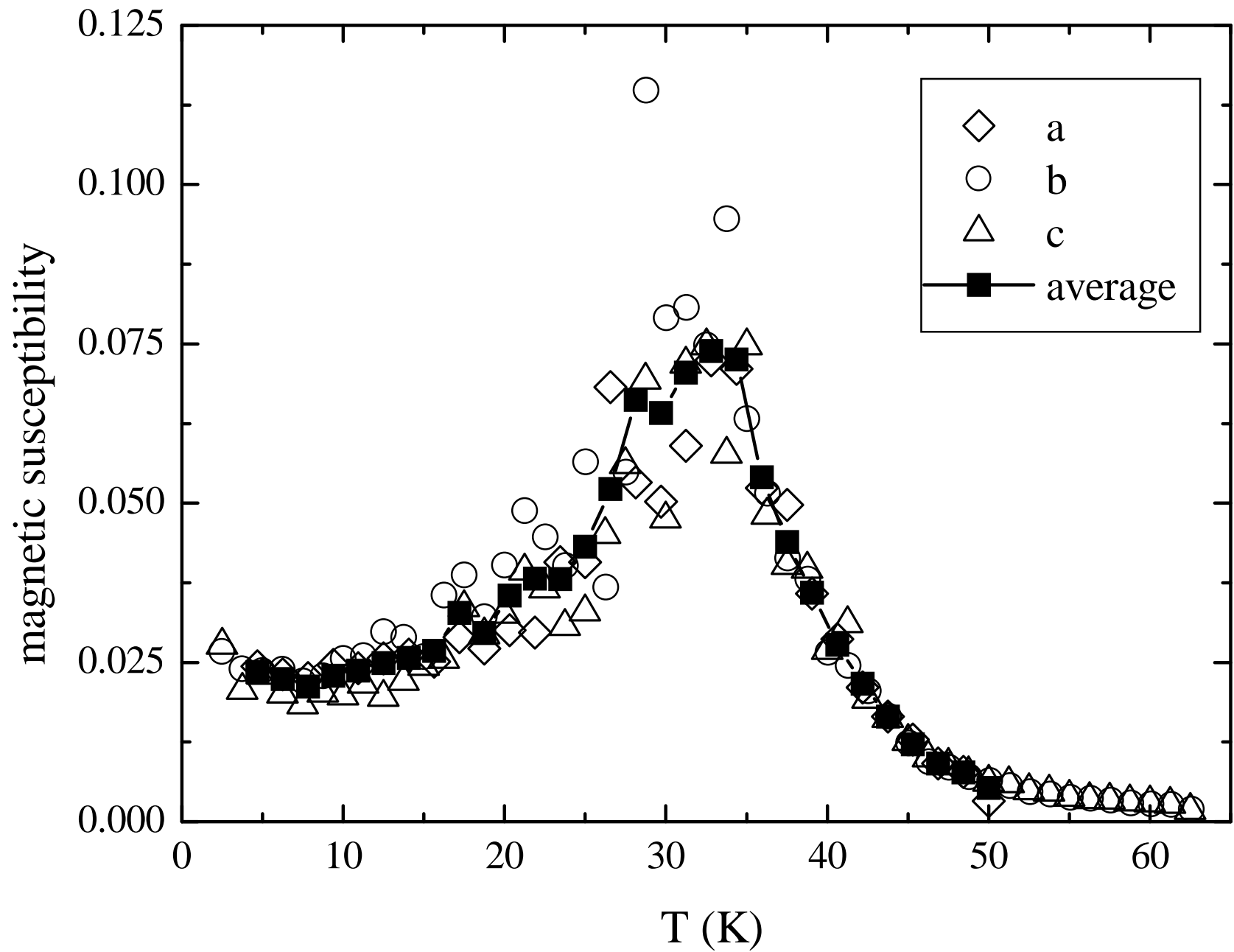
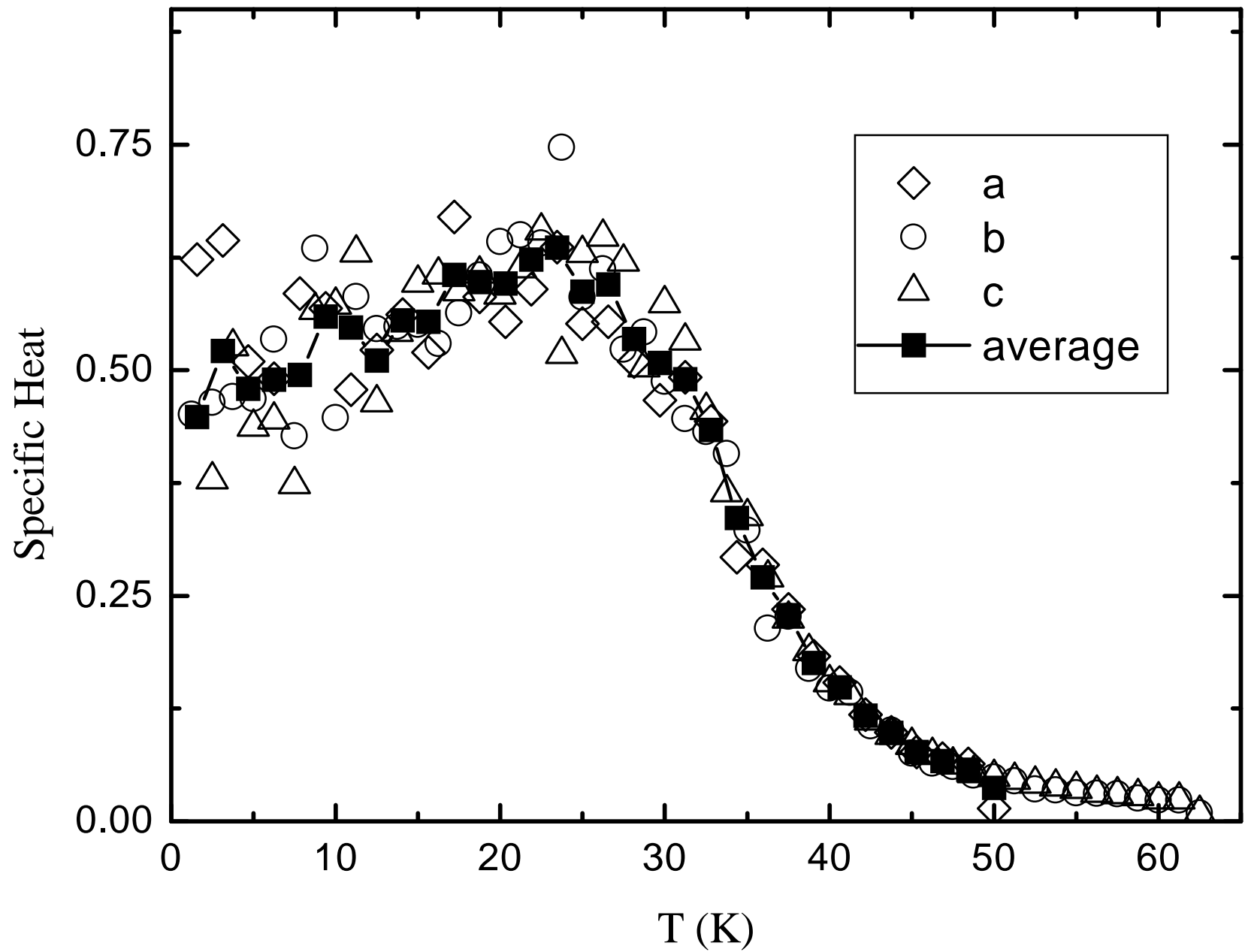
\end{document}